\documentclass[english,prl, twocolumn]{revtex4}
\usepackage[T1]{fontenc}
\usepackage[latin9]{inputenc}
\usepackage{graphicx}
\usepackage{babel}
\usepackage{amsmath, color}

\newcommand{\bccrit}{\ensuremath{(b/c)^\star}}
\newcommand{\fg}{\ensuremath{f_{\text{g}}}}
\newcommand{\fc}{\ensuremath{f_{\text{C}}}}
\newcommand{\fd}{\ensuremath{f_{\text{D}}}}
\newcommand{\SGE}{selective group extinction}

\begin{document}

\title{
 Promotion of Cooperation by Selective Group Extinction
}

\author{Marvin B\"ottcher$^{1,2}$ and Jan Nagler$^{3}$}

\affiliation{$^1$Department for Evolutionary Theory, Max Planck Institute for Evolutionary Biology, Pl\"on, Germany}
\affiliation{$^2$Max Planck Institute for Dynamics and Self-Organization, G\"ottingen, Germany}
\affiliation{$^3$Computational Physics, IfB, ETH Zurich, Wolfgang-Pauli-Strasse 27, 8093 Zurich, Switzerland}

\begin{abstract}
Multilevel selection is an important organizing principle that crucially underlies evolutionary processes from the emergence of cells to eusociality and the economics of nations. 
Previous studies on multilevel selection assumed that the effective higher-level selection emerges from lower-level reproduction.
This leads to selection among groups, although only individuals reproduce.
We introduce {\em \SGE}, where groups die with a probability inversely proportional to their group fitness.
When accounting for this the critical benefit-to-cost ratio is substantially lowered.
Because in game theory and evolutionary dynamics the degree of cooperation crucially depends on this ratio above which cooperation emerges previous studies may have substantially underestimated the establishment and maintenance  of cooperation.

\end{abstract}

\maketitle


The evolution of cooperation within and between species has been a puzzling phenomenon since the early beginnings of the theory of evolution.
Despite the competition between individuals and their selfish struggle for survival, cooperation between individuals emerged in all areas of life, which is sometimes termed ``the problem of cooperation'' \cite{Axelrod1985evolution}.

The problem of cooperation is commonly described in the framework of evolutionary game theory where individuals participate in a game whose outcome influences their success in the population over time \cite{Nowak2006evolutionary}. 
In these games individuals have the choice to either cooperate or defect.
This choice then includes a \emph{social dilemma}: defectors always win against cooperators, but the total payoff for the players is higher if they both cooperate.
A variety of mechanisms were proposed that contribute to the evolution of cooperation in the context of this social dilemma \cite{Axelrod1985evolution,Nowak2006five}. 
A strong promoter for cooperation can be the structure of a population itself, allowing for the emergence of clusters of cooperators \cite{Nowak1992,KillingbackDoebeli1996spatial,Traulsen2006evolution, Szabo2007, Santos2008diversity}, 
including the potentially beneficial effects for cooperation if the population structure changes over time \cite{Perc2010}.

The idea that competition between groups can lead to the establishment of cooperation under evolutionary selection pressures can be traced back to Darwin, who argued that group selection will occur when the benefits of cooperation or altruism between subpopulations are greater than the individual benefits of egoism within a subpopulation \cite{Darwin1871descent}.
In fact, the success of cooperation on the group level seems to be witnessed at all levels of biological or social-economical organization.
This applies especially in the context of humans which show high cooperation towards members of the same (political, religious or other) group and are strongly adverse to different groups \cite{Choi2007coevolution}.
Following Maynard Smith, group selection {\em requires that groups be able to ''reproduce'', by splitting or by sending out propagules, and that groups should go extinct} \cite{Smith1976group}.
An early mathematical foundation for multilevel selection was laid by Wright in 1945 \cite{Wright1945tempo}. 

Evolutionary theory suggests, however, that for a whole group to get a single trait, it must spread through the whole group first by regular evolution. 
Therefore group and kin selection are not always independent concepts \cite{Dawkins1979twelve, Dugatkin1994behavioral, Reeve2007emergence, Wenseleers2010social, Lehmann2007group}, and kin selection \cite{Hamilton1964genetical, Dawkins1979twelve} or reciprocation \cite{Nowak2006five} (or other concepts different from group selection) can dominate the evolution of cooperation \cite{West2007evolutionary}.
Whether or not, and when, kin and group selection are equivalent concepts crucially depends on the assumptions made and even on the mathematical description \cite{Wilson1975theory, Wilson1997, Crow1984group, Gardner2007relation, Lehmann2007group, West2007social, Wilson2007} which both have been controversially debated in past years \cite{Traulsen2010mathematics,Nowak2010evolution, Bourke2011, Veelen2012group}.

In this paper we introduce {\SGE} and study the consequences for the establishment and maintenance of cooperation in evolutionary dynamics for a minimal set of assumptions.
Specifically, {\SGE} might play an important role in situation of direct competition between groups like war, famines or other events where the survival of the whole group is at stake \cite{Choi2007coevolution}. 

Specifically, we study {\SGE} based on a simple stochastic model for group selection by Traulsen and Nowak \cite{Traulsen2006evolution} (referred to as the TN model thereafter).
The TN model is based on the assumption that groups die independently of the fraction of cooperators in the group and it enables the evolution of altruism in populations where the individuals compete in a prisoners dilemma game but without reciprocation or kin selection.
The advantages that cooperating groups are expected to have by utilizing coordinated efforts to outcompete groups of selfish individuals may not be ignored. 
This calls for a certain direct form of selective group extinction, independently of the absence - or presence - of other important mechanisms that promote cooperation in structured populations by direct competition \cite{Perc2013}.
In particular, since cooperator groups emerge and persist due to advantageous intra-group cooperation and get necessarily evolutionary unstable when cooperation is lost, we drop the assumption of fitness-independent group extinction \cite{Tarnita2009, Nowak2010, Perc2013}.

So we suggest {\SGE}, where groups die with a probability inversely proportional to their group fitness.
Specifically, we assume the group fitness to be a linear function of the fraction of cooperators in the group, which enables us to quantify the effect analytically for certain limiting cases.
Perhaps most importantly, we 
demonstrate a rather substantial impact 
of {\SGE} on the critical benefit-to-cost ratio above which cooperation emerges.

\section{Model}

 \begin{figure}[t] 
\includegraphics[width=0.45\textwidth]{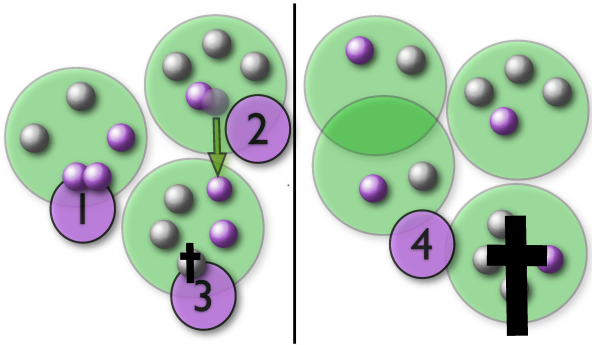}
\caption{
\label{fig:sketch} 
(Color online)
{\bf Model.}
The whole population of individuals is divided into $M$ groups (here $M=3$).
Individuals are reproduced according to a fitness which is calculated from the payoff in an evolutionary game between the individuals in each group. 
One individual is chosen stochastically according to its fitness and reproduces; the offspring is placed in the same group with probability $1-\lambda$ (event 1), or migrates with probability $\lambda$ to a randomly chosen group (2).
If a group exceeds the maximum size $N$, either one randomly chosen individual of the group dies with probability $1-q$ (3) or the group divides with probability $q$ (4).
At a group splitting event (4), all individuals are randomly distributed into two groups and one group is
chosen for elimination with probability inversely proportional to the group fitness $\fg$.
}
\end{figure}

The prisoners dilemma is played between two individuals who have a choice between cooperate (C) or defect (D): 
If both players cooperate they get the \emph{reward} $R$; if both defect, they receive the \emph{punishment} payoff $P$;
If one of the players defects while the other cooperates, the defecting player gets the \emph{temptation} payoff $T$ and the other player gets the \emph{sucker's} payoff $S$. 
The classical prisoners dilemma requires $T>R>P>S$.
Rational choice thus leads the two players to both play defect even though each player's individual reward would be greater if they both played cooperate.
In equilibrium, each individual chooses to defect even though both would be better off by cooperating, hence the dilemma.

To further stress the problem of cooperation, we use a common alternative parametrization for the prisoner's dilemma: cooperators pay the \emph{cost} $c$ to generate the \emph{benefit} $b$ for the other player, while defectors do not pay a cost and do not generate a benefit.
Thus the game is characterized by the payoff matrix for player A
\begin{align*}
 \bordermatrix{
    \frac{\text{player B}\rightarrow}{\text{player A}\downarrow}& C & D \cr
    \hspace{0.5cm}C & R=b-c & S=-c \cr
    \hspace{0.5cm}D & T=b & P=0 \cr
}
\end{align*}

In order to achieve analytic insights, we use the standard parametrization of the Prisoner's Dilemma (other choices do not change any of the main results).
We assume that individuals are either C or D and do not change their behavior over their life span. 
If they reproduce, the offspring has the same behavior as the ancestor which means that individuals reproduce via cloning.
The whole population is divided into $M$ groups with a group size limit $N$.
The individuals compete in their respective groups via the prisoners dilemma game.
The evolution of the population is determined
by the following events (Fig.\ \ref{fig:sketch}):

{\em Reproduction:} According to its fitness one individual is randomly chosen to 
produce an offspring. The fitness of a cooperator or defector in a group of $i$ cooperators and $n-i$ defectors with selection strength $w$ ($0\le w\le 1$) is given by \cite{Nowak2004emergence}:
\begin{equation}
\begin{aligned}
\fc(i) &= 1 + w - w \frac{R~(i-1)+S~(n-i)}{n-1}\\ \label{eq:TNfitness}
\fd(i) &= 1 + w - w \frac{T~i+P(n-i-1)}{n-1} 
\end{aligned}
\end{equation}
The offspring is placed in the same group (with probability $1-\lambda$) as the ancestor and shares the same strategy.
{\em Migration:} With probability  $\lambda$ the offspring of the reproduced individual migrates into another group which is also randomly chosen.\\
{\em Group splitting:} If the group size exceeds a given threshold $N$ via reproduction or migration,
 one individual in this group will \emph{die} with probability $1-q$, or with probability $q$ the group splits up. 
All individuals of the splitting group are randomly
 divided into two new groups. No empty groups are allowed. To keep the number of groups in the population constant,
 one group is chosen for elimination afterwards.\\
{\em Selective group extinction:}
The doomed group is chosen with probability inversely proportional to its group fitness
\begin{equation}\label{eq:pg}
p_{\text{g}}^\dagger=f_{\text{g}}^{-1} / \sum_j f_{\text{g}_j}^{-1}
\end{equation}
where the index $j$ runs over all groups and we choose the \emph{group fitness} $\fg$ to be linearly dependent on the fraction of cooperators in the group,
\begin{equation}\label{eq:fg}
\fg=\frac{i~\fc(i)+(n-i)~ \fd(i)}{n}.
\end{equation}
Definition Eq.(\ref{eq:fg}) fulfills the following requirements: (i) $\fg$ is a function of both the fraction of cooperators in the group $i/n$ and the fraction of defectors in the group $(n-i)/n$,
(ii) $\fg=\fc$ for groups with only cooperators,
(iii) $\fg=\fd$ for groups with only defectors,
and (iv) the group fitness $\fg$ is dependent on the selection strength $w$. 
This ensures that for $w=0$ the fitness is identical for all groups, whereas for $w=1$ a maximum fitness difference between homogeneous cooperator groups 
and homogeneous defector groups is guaranteed.

\section{Results}

\begin{figure} 
    \includegraphics{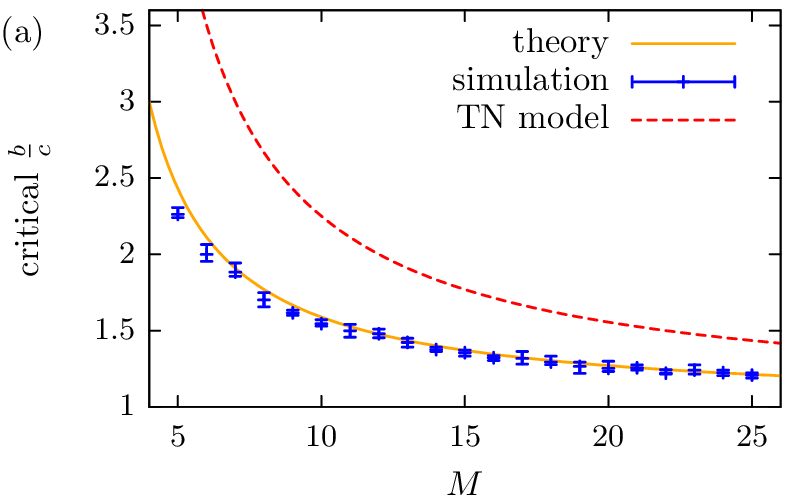}\\ \vspace{0.3cm}
\includegraphics{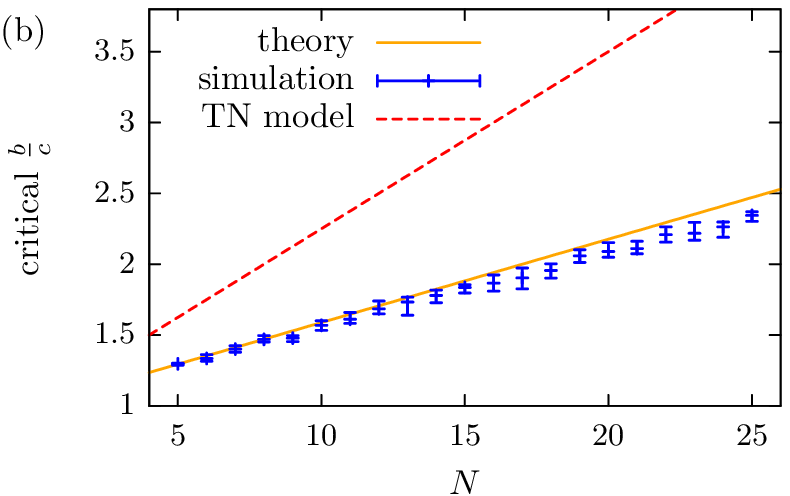}
\caption{\label{fig:results}(Color online)
{\bf Impact of group number and group size on $\mathbf{\bccrit}$.}
Comparison of the critical benefit-to-cost ratio $\bccrit$ of our model with the model of Traulsen and Nowak \cite{Traulsen2006evolution}.
In (a) the group size is fixed at $N=10$ and in (b) the number of groups is $M=10$.
Numerical simulations with group splitting probability $q=10^{-3}$ and selection strength $w=0.1$ agree perfectly with the theoretical results of Eq.\ (\ref{eq:bcComplete}).
For both (a) and (b) our model exhibits a substantially lower critical benefit-to-cost ratio and hence a larger benefit for cooperators.}
\end{figure}

\begin{figure} 
\includegraphics{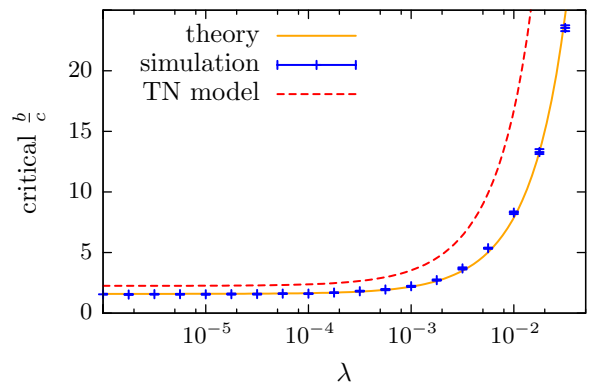}
\caption{\label{fig:results2}(Color online)
{\bf Impact of migration on $\bccrit$.}
Comparison of the critical benefit-to-cost ratio $\bccrit$ against the migration rate $\lambda$ of our model with the TN model.
Numerical simulations with group splitting probability $q=10^{-3}$ and selection strength $w=0.1$ agree perfectly with the theoretical results of Eq.\ (\ref{eq:bcComplete}).
Our model exhibits a higher robustness against migration.
}
\end{figure}

To measure for which parameter regimes cooperation can emerge, we determine the \emph{critical benefit-to-cost ratio} $\bccrit$ by comparing the fixation probability of one cooperator in the population of defectors $\rho_C$ and the fixation probability of one defector in the population of cooperators $\rho_D$.
At $b/c=\bccrit$ both strategies have the same fixation probability.
If the benefit-to-cost ratio is above this critical limit, cooperation is advantageous, while below this limit defection is the profitable strategy.

Consideration of certain limiting cases allow for analytical results \cite{Traulsen2006evolution}.
In particular, we consider a low group splitting probability $q\ll 1$ and small migration rate $\lambda\ll 1$.
This leads to a separation of timescales between the processes inside a single group and the group splitting events between different groups.
Therefore our model becomes a two-level Moran process \cite{Moran1958random, Moran1962statistical} with fixation of single individuals in a given group in the first level and then fixation of a homogeneous group among all other homogeneous groups on the second level.
Note that for similar models, but without group splitting, the requirement for small migration probabilities can be relaxed while still allowing for analytical results \cite{Antal2009, Fu2013globalmigration, Fu2012favoritism}.
However, these models do not include direct competition between groups or a certain group fitness.

Given the update rules, we calculate the transition probabilities to jump from $j$ to $j+1$ or from $j$ to $j-1$ pure cooperator groups, respectively.
For the jump from $j$ to $j+1$ cooperator groups it is calculated by the probability that a cooperator group splits (dependent on single individuals fitness) multiplied with the probability that a defector group dies (dependent on group fitness) plus the probability that the offspring of a cooperator group migrates and takes over a defector group.
 \begin{eqnarray}\label{eq:trans}
 \begin{aligned}
   P_{j,j+1}   = \frac{f_C(N)j}{(f_C(N)j+f_D(0)f(M-j))} \times\\
         \left(q\frac{f_C(N)(M-j)}{(f_D(0)j+f_C(N)(M-j))}+\lambda \varphi_C \frac{(M-j)}{M}\right) \\ 
   P_{j,j-1} = \frac{f_D(0)(N-j)}{(f_C(N)j+f_D(0)(M-j))} \times \\
         \left(q\frac{f_D(0)j}{(f_D(0)j+f_C(N)(M-j))}+\lambda \varphi_D \frac{j}{M}\right) 
  \end{aligned}
 \end{eqnarray}
Here the variables $\varphi_C$ and $\varphi_D$ are the fixation probabilities for C and D individuals in a single group and are identical to Ref.\ \cite{Traulsen2006evolution}:
 \begin{equation}
 \begin{aligned}
    \varphi_C &=\frac{1}{N}\left[1-\frac{w}{6}\delta_C\right]\\
   \varphi_D &= \frac{1}{N}\left[1-\frac{w}{6}\delta_D\right]
 \end{aligned}
 \end{equation}
 with $\delta_C := \left[(T-R+2P-2S)N+(T-4R+2P+S)\right]$ and $\delta_D :=\left[(-2T+2R-P+S)N-(T+2R-4P+S)\right]$.

From the transition probabilities, Eq.\ (\ref{eq:trans}), we derive the fixation probability for cooperator groups \cite{Moran1962statistical, Traulsen2006evolution},
 \begin{align}
   \Phi_C   = \left[ 1+ \sum_{k=1}^{M-1}\prod_{j=1}^k \frac{P_{j,j-1}}{  P_{j,j+1}} \right]^{-1}
 \end{align}
 For small selection strength $w$ we can Taylor expand the expression for group fixation at $w=0$ \footnote{The assumption $w\ll 1$ can be dropped by introducing an exponential fitness function in dependence on the payoff of the prisoners dilemma game \cite{Traulsen2008analytical}} and then arrive at an analytical expression for both $\Phi_C$ and $\Phi_D$:
\begin{align*}
\Phi_C = \frac{1}{M}\left[1-w\frac{M-1}{2} \Gamma\right], \;\;
  \Phi_D = \frac{1}{M}\left[1+w\frac{M-1}{2} \Gamma\right]
\end{align*}
with
\begin{align*}
        \Gamma := P-R+\frac{N}{Nq+\lambda}\left(-q(R-P)+\frac{\lambda}{6N}(\delta_C-\delta_D)\right).
 \end{align*}

From small group splitting probability $q\ll 1$ and small migration probability $\lambda\ll 1$ we can assume that in case of a group splitting event the individual groups all have size $N$ and are either completely cooperating or defecting. 
To get the overall fixation probabilities $\rho_C$ and $\rho_D$ for single individuals in the whole population we therefore take the product of the fixation probability of one individual in the group and the fixation probability of this cooperator or defector group in the whole population, $\rho_C=\varphi_C \Phi_C$ and $\rho_D=\varphi_D \Phi_D$.

Cooperation is advantageous, if $\rho_C$ is larger than $\rho_D$ with the critical benefit-to-cost limit given by $\rho_C=\rho_D$. 
For arbitrary $N$ and $M$, but under the given constraints of small group splitting probability $q$, low migration probability $\lambda$ and small selection strength $w$, we eventually derive the critical benefit-to-cost ratio,
\begin{equation}\label{eq:bcComplete}
(b/c)^\star = 1+ \frac{N/M + \frac{\lambda}{q}}{(2-3/M)-\frac{\lambda}{q}/(NM)}.
\end{equation}

For zero migration, $\lambda=0$, which corresponds to perfect separation of groups and no noise in strategies, Eq.\ (\ref{eq:bcComplete}) reduces to to the simpler
$(b/c)^\star = 1+\frac{N/M}{(2-3/M)}$.
For the asymptotics $N\gg 1$ and $M\gg 1$ (and $\lambda$ constant) we obtain
\begin{equation}\label{eq:bcasym}
(b/c)^\star =1+\frac{N/M + \frac{\lambda}{q}}{2}.
\end{equation}

We have tested our analytical predictions for the critical benefit-to-cost ratio by numerically determining the point where from evenly mixed population cooperators and defectors take over the whole population with the same frequency. 
Specifically, we used a nested interval method to find the critical benefit-to-cost ratio for each parameter set of $N$, $M$ and $\lambda$.
Our simulations are in excellent agreement with the theory (Figs.\ \ref{fig:results} \& \ref{fig:results2}; error bars give the first and the third quartile, respectively). 

\section{Discussion}
The combined selection at birth and death of groups \cite{Altrock2009deterministic} considerably favors cooperation, as it promotes the selective advantage of homogeneous groups of cooperators. 
This results in a significantly lowered critical limit of the benefit-to-cost ratio $\bccrit$ as a function of $N$, $M$ (Fig.\ \ref{fig:results})  and $\lambda$ (Fig.\ \ref{fig:results2}) above which cooperation is advantageous.

Similarly to the results of Traulsen and Nowak \cite{Traulsen2006evolution} the critical benefit to cost ratio increases linearly with groups size $N$ and decreases inversely to the number of groups $M$ (Fig. \ref{fig:results}).
However, the critical ratio of benefit over cost, above which cooperation emerges and is maintained in the population, is significantly lower; in the limit of large group sizes $N$ the critical ratio $\bccrit$ is approximately half.
This means that with {\SGE} cooperation is advantageous for group sizes twice as large as in the TN model or for half the number of groups. 
Correspondingly, with the same population structure $N$ and $M$ a much lower $\bccrit$ can be chosen, allowing cooperation to emerge for a broader range of altruistic acts. 

The population size 
can  play a decisive role in evolutionary dynamics \cite{Nowak2004emergence, Fudenberg2006}.
In our model, the difference in the critical benefit-over-cost ratio $\bccrit$ between our model and the TN model \cite{Traulsen2006evolution} behaves differently for the varying number of groups $M$ or varying group size $N$.
For no migration the difference is
\begin{align}
    \Delta_{\bccrit}= \frac{N(M-1)}{(M-2)(2M-3)} \propto \frac{N}{M}
\end{align}
which effectively increases linearly with $N$ and is inversely related to $M$.
For increasing number of groups $M$ the difference between the TN model and {\SGE} decreases due to similar extinction probabilities of homogeneous groups after group splitting events:
If there is only a single homogeneous cooperator group among many defector groups, one of the defector groups will die with a high probability anyway, even without the effect of higher fitness of the single cooperator group with {\SGE}, whereas for for a low number of groups {\SGE} protects this cooperator group from extinction. 
For increasing group size $N$, on the other hand, the difference in $\bccrit$ increases since with {\SGE} the fitness of defector groups is greatly reduced.
For large group sizes defectors overtake a single group with a much higher probability than cooperators, resulting in the steep linear increase of the critical benefit-to-cost ratio with group size $N$.
However, for defectors to fixate in the population, they also need to spread through the population via group splitting events; with {\SGE} this is much harder as homogeneous defector groups are more likely to die which greatly diminishes the benefit of large groups for defectors.

Again similar to the TN model, successful evolution of cooperation requires a low probability of migration $\lambda/q$ (cf.\ Eq.\ \ref{eq:bcasym}) and hence strongly separated groups. 
In particular, given a fixed group splitting probability $q$, $\bccrit$ diverges at $\lambda_d=qN(2M-3)$ (Fig.\ \ref{fig:results2}) and reaches already twice the $\bccrit$ as without migration if $\lambda_{2\times}\approx q\times 1.664$ for both $N$ and $M$ equal to ten.
Therefore it is advantageous to hinder the  invasion of individuals into groups and thus prevent group mixing.
Generally, group selection models are highly vulnerable against migration of defecting individuals \cite{Traulsen2006evolution}, as single defectors can easily overtake cooperator groups.
This is also true for other structured population models \cite{Szolnoki2009,Szolnoki2009resolving}: if the number of random links between individuals becomes too large, the benefit for cooperation decreases, even though some mixing is always required to allow for successful fixation in a population.
Accordingly also the different forms of migration can be crucial to the evolutionary dynamics in the population \cite{Pichugin2015modesofmigration}.
If for very high migration rates the population structure vanishes and the population becomes well-mixed, defection is the favorable strategy according to the social dilemma \cite{DongPing2009effect}. 

However, selective group extinction greatly increases the robustness of the degree of cooperation against the migration of individuals, which can be interpreted as robustness against mixing or diffusive noise: the migration rate which leads to diverging $\bccrit$ is smaller by a factor of approximately two $\lambda_d=qN(M-2)$.
Increased robustness of cooperation in conditions of increased mixing can therefore be a noticeable advantage for the evolution of cooperation with \SGE.

Several factors not included into the current model might influence the effect of selective group extinction:
Firstly, a non-linear group fitness function can have a strong impact on the fixation probability.
Our analytical results do not depend on the particular shape of the group fitness function which we chose to be linear, since in case of a group splitting event we assume all groups to be homogeneous. 
For higher group splitting probabilities, however, the particular shape of the group fitness function does matter and  a plethora of interesting effects might be observable dependent on the particular choice of the fitness function \cite{Szolnoki2010,Archetti2012review}.

Secondly a variable size of groups might have beneficial effects on the evolution of cooperation by group selection. Groups with more cooperators will grow faster which in turn increases the share of cooperators in the population and increases the frequency pure cooperator groups after group splitting.

\section{Conclusion}
In conclusion, direct selection between groups via {\SGE} substantially lowers the critical cost-to-benefit ratio $\bccrit$ both as a function of the group size $N$, the number of groups $M$ and the migration probability $\lambda$.
In game theory and evolutionary dynamics the establishment and maintenance  of cooperation 
 is generically a threshold phenomenon \cite{Hauert2005, Nowak2006five}. Thus, any substantial
 change in the threshold is expected to have rather a drastic effect.
For instance, given a fixed benefit $b$, a change of the cost $c$ of 20\% usually crucially determines the establishment and maintenance  of cooperation, both in simple and more complex social and economic systems \cite{Neumann1944, Axelrod1985evolution, Kagel1995, Hauert2005}.
In our model, depending on the circumstances and compared with the TN model, the lowering of $\bccrit$ is in the range of 50 to 80$\%$, which constitutes our main result.
Thus, our results indicate that {\SGE} {\em substantially} promotes the advantageous effects of group selection for cooperation.
Selective group extinction might contribute to group selection in social behavior of animals or plants \cite{Goodnight1997experimental}, cultural and economic behavior of humans \cite{Henrich2004cultural} or even cooperative effects in the microbial world \cite{Velicer2003social}.

 \paragraph{Acknowledgments}
Jan Nagler thanks the ETH Risk Center for support.

\bibliography{references}

\end{document}